\documentstyle[prb,aps,twocolumn,epsf]{revtex}

\def\prb{\it Phys. Rev. B}
\def\prl{\it Phys. Rev. Lett.}

\def\be{\begin{equation}}
\def\ee{\end{equation}}
\def\ba{\begin{eqnarray}}
\def\ea{\end{eqnarray}}

\def\C60{A$_x$C$_{60}$}

\def\hty{high temperature superconductivity}
\def\hts{high temperature superconductors}
\def\htr{high temperature superconductor}

\makeatletter
\def\@oddhead{{\hfill \small\sf To appear as a Perspective in {\it Proc.\ Natl.\
Acad.\ Sci.\ USA}}}
\let\@evenhead\@oddhead
\makeatother

\begin{document}

\twocolumn[\hsize\textwidth\columnwidth\hsize\csname@twocolumnfalse\endcsname

\title
{Stripe Phases in High Temperature Superconductors}

\author{ V.~J.~Emery$^1$, S.~A.~Kivelson$^2$, and J.~M.~Tranquada$^{1}$ }
\address{
1) Dept. of Physics,
Brookhaven National Laboratory,
Upton, NY  11973-5000}
\address
{2)  Dept. of Physics,
U.C.L.A.,
Los Angeles, CA  90095}
\date{\today}
\maketitle

\widetext

\begin{abstract}
Stripe phases are predicted and observed to occur in a class of
strongly-correlated materials describable as doped
antiferromagnets, of which the copper-oxide superconductors are the most
prominent representative.  The existence of stripe correlations necessitates
the development of new principles for describing charge transport, and
especially superconductivity, in these materials.
\end{abstract}

\vspace{0.2in}

]
\makeatletter
\global\@specialpagefalse
\makeatother

\narrowtext

Thirteen years ago, the discovery\cite{bedn86} of superconductivity in layered
copper-oxide compounds came as a great surprise, not only because of the
record-high transition temperatures, but also because these materials are
relatively poor conductors in the  ``normal'' ({\it i.e.}, nonsuperconducting)
state. Indeed, these superconductors are obtained by electronically doping
``parent'' compounds that are antiferromagnetic Mott insulators, materials in
which both the antiferromagnetism and the insulating behavior are the result of
strong electron-electron interactions.  Since local magnetic correlations
survive in the metallic compounds, it is  necessary to view these materials as
doped antiferromagnets. A number of other  related materials, such as the layered
nickelates (which remain insulating when doped) and manganites (the ``colossal''
magnetoresistance materials),  are also doped antiferromagnets, in this sense.

The conventional quantum theory of the electronic structure of
solids,\cite{history} which has been outstandingly successful at describing the
properties of good electrical conductors (metals such as Cu and Al) and
semiconductors (such as Si and Ge), treats the electronic {\it excitations}
as a weakly interacting gas. This approach, known as ``Fermi Liquid Theory'',
breaks down when applied to doped
antiferromagnets.   New {\it principles} must be developed to deal with
these problems, which are at the core of the study of ``strongly correlated
electronic systems'', one of the central and most intellectually rich branches
of contemporary physics.   One idea that has evolved over the last decade, and
which offers a framework for interpreting a broad range of experimental
results on copper-oxide superconductors and related systems, is the concept of a
stripe phase. A stripe phase is one in which the doped charges are concentrated
along spontaneously generated domain walls between  antiferromagnetic insulating
regions.

Stripe phases occur  as a compromise  between the
antiferromagnetic interactions among magnetic ions and the Coulomb interactions
between charges (both of which favor localized electrons)
and the zero-point kinetic energy of the doped holes (which tends to delocalize
charge). Experimentally, stripe phases are most clearly detected in  insulating
materials (where the stripe order is relatively static), but there is
increasingly strong evidence of fluctuating stripe correlations in metallic and
superconducting compounds.   The existence of dynamic stripes, in turn, forces
one to consider new mechanisms for charge transport and for superconductivity.
More generally, we will show that the concept of electronic stripe phases
developed for transition-metal oxides is applicable to a broad range of
materials.

{\bf Theoretical Background.}
Doped antiferromagnets are a particularly important and well-studied class of
strongly-correlated electronic materials.   Here, the parent compound is
insulating, even at elevated temperatures,  because of the strong short-range
repulsion between electrons.  At sufficiently low temperatures,
antiferromagnetic order develops in which there is a non-zero average magnetic
moment on each site pointing in a direction that alternates from site to site.
(See Fig. 1.) Frequently the doping process, ``hole doping'', involves
chemically modifying the material  so that a small fraction of electrons is
removed from the insulating antiferromagnet. Whereas the charge distribution in
a doped semiconductor is homogeneous, in a doped antiferromagnet the added
charge forms clumps: solitons in one dimension, linear ``rivers of charge'' in
two dimensions, and planes of charge in three dimensions, as exemplified by
organic conductors, cuprates or nickelates, and manganites respectively.
Typically, these clumps  form what are known as ``topological defects'' across
which there is a change in the phase of the background spins or orbital degrees
of freedom. In $d$ dimensions, the defects are $(d-1)$-dimensional extended
objects.\cite{larged}
Stripes in a two-dimensional (2D) system are illustrated
schematically in Fig. 1.

Self-organized local inhomogeneities were predicted
theoretically.\cite{zaanen,schulz,physicaC,ute}  They arise because the
electrons tend to cluster in regions of suppressed antiferromagnetism\cite{bag}
which produces a strong, short-range tendency to phase
separation\cite{ekl1,ekl2,manousakis} that is frustrated by the long-range
Coulomb interaction. The best compromise\cite{ute,chayes} between these competing
imperatives is achieved by
allowing the doped holes to be delocalized along linear stripes,  while the
intervening regions remain more-or-less in the undoped correlated insulating
state.

{\bf Experimental Evidence for Stripes.}
The most direct evidence for stripe phases in doped antiferromagnets has come
from neutron scattering studies.  Diffraction of a neutron beam by long-period
spin and charge density modulations, extending over a few unit cells,
as indicated in Fig.~1, yields
extra Bragg peaks.  The position of such a superstructure peak measures the
spatial period and orientation of the corresponding density modulation, while
the intensity provides a measure of the modulation amplitude.  Since neutrons
have no charge, they do not scatter directly from the modulated electron
density, but instead are scattered by the ionic displacements induced by
the charge modulation.  The lattice modulation is also measurable with electron
and x-ray diffraction.

\begin{figure}[t]
\epsfxsize=3.3in
\centerline{\epsfbox{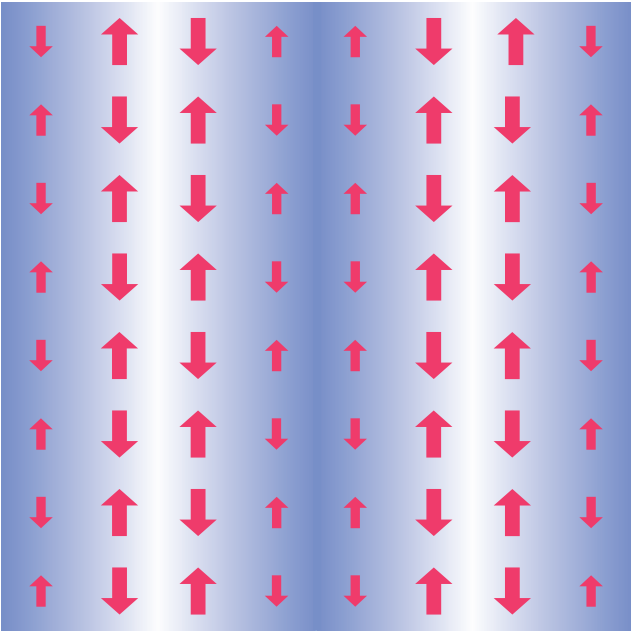}\medskip}
\caption{
Schematic picture of a stripe ordered phase.  The arrows represent the
magnetic or spin order, and the blue scale represents the local charge density.
Regions of high charge density (stripes) lie between largely undoped
regions, where the spin order is much the same as in the undoped
antiferromagnet.  In the figure, the stripes lie along the direction
of the nearest-neighbor bonds, which we refer to as ``vertical'' stripes;
when the stripes lie at 45$^{\circ}$ to this axis they are said to
be ``diagonal.''}
\label{Fig1}
\end{figure}

The antiferromagnetic order found in the parent compounds of the cuprate
superconductors is destroyed rapidly as holes are introduced by doping. The
first indications of long-period (``incommensurate'') spin-density modulations
were provided by inelastic neutron scattering studies\cite{cheo91} of
superconducting La$_{2-x}$Sr$_x$CuO$_4$, and by related measurements on the
insulating nickelate analog.\cite{hayd92}  Following the discovery of
``incommensurate'' charge ordering in the latter system by electron
diffraction,\cite{chen93} the proper connection between the magnetic and
charge-order peaks was determined in a neutron diffraction study\cite{tran94a}
of La$_2$NiO$_{4.125}$.  The positions of the observed peaks indicate that the
charge stripes run diagonally through the NiO$_2$ layers (as opposed to the
vertical stripes shown in Fig.~1).  More recent experiments\cite{lee97,yosh99}
on La$_{2-x}$Sr$_x$NiO$_4$ have shown that the diagonal stripe ordering  occurs
for doping levels up to $x\sim\frac12$ (corresponding to a hole density of 1
for every 2 Ni sites), with the maximum ordering temperatures occurring at
$x=\frac13$.

It is significant that the charge ordering is always observed at a
higher temperature than the magnetic ordering, which is
characteristic\cite{zachar} of a transition that is driven by the
charge. It is also important
to note that the period  of the charge order
is generally temperature dependent, which means that the hole concentration
along each stripe also varies with temperature;  this behavior is
characteristic\cite{john} of
structures that arise from competing interactions. 
These observations are consistent with the idea that the stripes are generated
by the competition between the clustering tendency of the
holes and the long-range Coulomb interactions.
[Weak density-wave order can occur in conventional solids under
special conditions (``nested Fermi surfaces"), but the transitions
tend to be ``spin driven'' and occur at a fixed ``nesting'' wave
vector.\cite{schulz}]

Charge order is most easily detected when stripes are static, but  perfect
static charge order can be shown\cite{fradkin}  to be incompatible with the
metallic behavior of the cuprates. Nevertheless, to get a better
experimental handle on the charge order, one might hope to pin down fluctuating
stripes with a suitably anisotropic distortion of the crystal structure.  Just
such a distortion of the La$_{2-x}$Sr$_x$CuO$_4$ structure is obtained by
partial substitution of Nd for La.  Neutron diffraction
measurements\cite{tran95a} on a Nd-doped crystal with the special Sr
concentration of $x\approx\frac18$  revealed charge and spin order, consistent
with the vertical stripes of Fig.~1.  (An anomalous suppression of
superconductivity, associated with the lattice distortion, is maximum  for
$x\approx\frac18$.) The charge order has since been confirmed by high-energy
x-ray diffraction.\cite{vonz98}  As in the nickelates, the spin ordering
occurs at lower temperatures than the charge order, and the hole  concentration
on a stripe varies as a function of the Sr concentration, $x$.

Although it has been difficult to observe a direct signature of charge stripes
in other cuprate families, the existing neutron scattering studies of magnetic
correlations are certainly most easily understood in terms of the stripe-phase
concept.   The doping dependence of dynamic magnetic correlations\cite{yama98a}
in Nd-free La$_{2-x}$Sr$_x$CuO$_4$ is found to be essentially the same as the
static correlations in Nd-doped samples,\cite{tran95a} and a comprehensive
study\cite{aepp97} of a Nd-free sample near ``optimum'' doping ({\it i.e.},
maximum superconducting transition temperature) indicates that ordering may be
prevented by quantum fluctuations.  To keep things interesting, static magnetic
order has been observed\cite{lee99} to set in near the superconducting
transition temperature in La$_2$CuO$_{4+\delta}$.  Finally, a beautiful
experiment\cite{mook98} on superconducting YBa$_2$Cu$_3$O$_{6+x}$ has shown
that the low energy magnetic correlations in that system have strong
similarities with those in La$_{2-x}$Sr$_x$CuO$_4$.

An example of planar domain walls in a 3D system occurs in nearly-cubic
La$_{1-x}$Ca$_x$MnO$_3$ with $x\ge0.5$.  Charge order has been imaged by
transmission electron microscopy.\cite{mori98}  The ordering phenomena are
somewhat more complex in this case because the occupied Mn $3d$
orbitals are degenerate.  As a consequence, charge, spin, and orbital ordering
are all involved, although, again, charge order sets in at a higher temperature
than magnetic order.

{\bf Electronic Liquid Crystals.}
Once the idea of stripe phases of a two-dimensional doped insulator has been
established, a major question arises: How can a stripe phase become a {\htr},
as in the cuprates, rather than an insulator, as in the nickelates? 
Typically, interactions drive quasi one-dimensional metals to an
insulating ordered charge density wave (CDW) state at low temperatures\cite{emery}
(and quenched disorder only enhances the insulating tendency). However, we have
shown\cite{fradkin} that the CDW instability is eliminated and
superconductivity is enhanced if the transverse stripe fluctuations have a
large enough amplitude. To satisfy this condition, the stripes could oscillate
in time or be static and meandering. They are then electronic (and quantum
mechanical)  analogues of classical liquid crystals and, as such, they
constitute new states of matter, which can be either {\hts} or two-dimensional
anisotropic unconventional metals.

Classical liquid crystals are phases that are intermediate between a liquid and
a solid, and spontaneously break the  symmetries of free space.  Electronic
liquid crystals are quantum analogues of these phases in which the ground state
is intermediate between a liquid, where quantum fluctuations are large, and a
crystal, where they are small. Because the electrons exist in a solid, it is
the symmetry of the host crystal that is spontaneously broken, rather than the
symmetry of free space. An electronic liquid crystal has the following phases:
({\it i}) a {\it liquid}, which breaks no spatial symmetries and, in the absence
of disorder, is a conductor or a superconductor; ({\it ii}) a {\it nematic}, or
anisotropic liquid, which breaks the {\it rotation} symmetry of the lattice and
has an axis of orientation; ({\it iii}) a {\it smectic}, which breaks
translational symmetry in one  direction and, otherwise is an electron liquid;
({\it iv}) an {\it insulator} with the character of an electronic solid or
glass.  These classifications  applied to stripe phases make the stripe notion,
which is based on  local electronic correlations, macroscopically precise.
Neutron and x-ray scattering experiments give direct
evidence of electronic liquid crystal phases (conducting stripe ordered phases)
in the cuprate superconductors.

{\bf Charge Transport.}
In the standard theory of solids,  the electron's kinetic energy is treated as
the largest energy in the problem, and the effects of electron-electron
interactions are introduced as an afterthought.  As a consequence, the
electronic states in normal solids are highly structured in momentum space
({\bf k}-space), and {\it therefore}, according to the uncertainty principle,
they are highly homogeneous in real space. Moreover,  as the ``normal''
(metallic) state is continuously connected to the ground state of the kinetic
energy, any phase transition to a low-temperature ordered phase is
necessarily\cite{pwa} driven by the potential energy, inasmuch as it involves a
gain in the interaction energy between electrons at a smaller cost of kinetic
energy.  For transport properties, the central concept of a mean free path $l$,
{\it i.e.} the distance an electron travels between collisions, is well defined
so long as $l$ is much larger than the  electron's de Broglie wavelength,
$\lambda_{F}$, at the Fermi energy.

A number of interesting synthetic metals discovered in the past few decades
seem to violate  the conventional theory. They are ``bad metals''
\cite{badmetal,tall} in the sense that their resistivities, $\rho(T)$, have a
metallic temperature dependence [$\rho(T)$ increases with the temperature $T$]
but the mean free path, inferred from the data by a conventional analysis, is
shorter than $\lambda_{F}$, so the concept of a state in momentum space would
be ill defined. Among the materials in question are the cuprate {\hts}; other
oxides including the ruthenates, the nickelates and the ``colossal
magnetoresistance materials'' (manganites); organic conductors; and
alkali-doped $C_{60}$. Most of these materials are doped correlated insulators,
in which the short-range repulsive interaction between electrons is the largest
energy in the system.  However, the ground state of this part of the
Hamiltonian is not unique, so the kinetic energy cannot simply be treated as a
perturbation;  such materials display substantial structure in both real space
and momentum space. As a consequence, the conventional theory\cite{history}
must be abandoned. Neither the kinetic energy nor the potential energy is
totally dominant, and both must be treated on an equal footing.

{\bf Superconductivity.}
The highly successful theory of superconductivity\cite{bcs} developed by
Bardeen, Cooper, and Schrieffer in the fifties was designed for good metals,
not for doped insulators.   A key issue, therefore, is the relation of stripes
to the mechanism of high temperature superconductivity. In fact, there is a strong 
empirical case for an intimate relation between these phenomena: ({\it i})
strongly condensed stripe order can suppress superconductivity (as it does in
La$_{1.6-y}$Nd$_{y}$Sr$_x$CuO$_4$), ({\it ii})  weak stripe ordering can, at
times, appear at the superconducting transition temperature $T_c$ (as it does in
La$_2$CuO$_{4+\delta}$), ({\it iii}) there is a simple, linear relation between
the inverse stripe spacing and the superconducting
$T_{c}$ observed in several materials\cite{yama98a,bala99} (including
La$_{2-x}$Sr$_x$CuO$_4$ and YBa$_2$Cu$_3$O$_{6+x}$),  and ({\it iv}) stripe
structure and other features of the doped insulator, together with high
temperature superconductivity, disappear as the materials emerge from the
doped-insulator regime (``over doping'').
Moreover, there is a clear indication that the
optimal situation for {\hty} is stripe correlations that are not too static or
strongly condensed, but also not too ethereal or wildly fluctuating. We have
argued\cite{ekz,fradkin,miami} that the driving force for the physics
of the doped insulator is the
reduction of the zero-point kinetic energy. This proceeds in three steps: ({\it
i}) the development of an array of metallic stripes lowers the kinetic energy
along a stripe, ({\it ii}) hopping of pairs of electrons perpendicular to a
stripe in the CuO$_2$ planes creates spin pairs on and in the immediate
neighborhood of a stripe, and ({\it iii}) at a lower temperature, pair hopping
between stripes creates the phase coherence that is essential for
superconductivity. Steps {\it ii} and {\it iii} lower the kinetic energy of
motion perpendicular to a stripe.

{\bf Generality of the Stripe Concept.}
The physics of charge clustering in doped correlated insulators is
general and robust,
so one might expect that local stripe structures would
appear in other related systems.
Indeed, topological doping has long been documented in the
case of quasi one dimensional charge-density-wave (CDW) systems, such as
polyacetylene;\cite{emery,heeger} it is an interesting open question whether it
occurs in other higher dimensional systems.  One recent, fascinating discovery
is the observation\cite{eisenstein,tsui} that, under appropriate circumstances,
quantum Hall systems ({\it i.e.} an ultra-clean 2D electron gas in a high
magnetic field)  spontaneously  develop a large transport anisotropy on cooling
below 150~mK.  It is likely that this anisotropy is related to stripe formation
on short length scales,\cite{shlovskii,chalker} and it apparently
reflects the
existence of an electronic nematic phase in this system.\cite{eduardo}

Stripe-like structures have also been observed (ref.~43 and references therein)
in many other systems with
competing interactions, on widely differing length scales.
Beyond this generality, the existence of spontaneously
generated local structures is clearly important for understanding all of the
electronic properties of synthetic metals, including the anomalous charge
transport and the mechanism of high temperature superconductivity. Many of
these implications have already been explored in considerable detail, but many
remain to be discovered. Here we content ourselves with a few general
observations.

The phenomena described above represent a form of ``dynamical dimension
reduction'' whereby, over a substantial range of temperatures and energies, a
synthetic metal will behave, electronically, as if it were of lower
dimensionality. This observation has profound implications because conventional
charge transport occurs in a high-dimensional state, and fluctuation effects are
systematically more important in lower dimensions.  In particular, in the
quasi-two dimensional high temperature superconductors, stripes provide a
mechanism for the appearance of quasi-one dimensional electronic physics, where
conventional transport theory fails, and is replaced by such key notions as
separation of charge and spin, and solitonic quasi-particles.\cite{emery}  At
the  highest temperatures (up to 1000~K), in what is often called the ``normal
state'' of the high temperature superconductors, where coherent stripe-like
structures are unlikely to occur, it is still probable that local charge
inhomogeneities occur due to the strong tendency of holes in an antiferromagnet
to phase  separate.\cite{losalamos} This behavior can lead to quasi
zero-dimensional physics (quantum impurity model
physics), which also produces a host of interesting, and well documented
quantum critical phenomena and may be at the heart of much of the anomalous
normal state behavior of these systems.

\end{document}